\documentclass[aps,prl,onecolumn,10pt,noshowpacs,noshowkeys, notitlepage,twoside,a4paper,nobibnotes]{revtex4-1}
\usepackage[usenames,dvipsnames]{color}
\usepackage{graphicx}
\usepackage{amsmath}

\def\be{\begin{equation}}
\def\ba{\begin{eqnarray}}
\def\ee#1{\label{#1}\end{equation}}
\def\ea#1{\label{#1}\end{eqnarray}}

\def\bs{\begin{center}}
\def\es{\end{center}}

\begin{document}
\title{Transient anomalous diffusion in periodic systems: \\ergodicity, symmetry breaking and velocity relaxation}
\author{Jakub Spiechowicz}
\affiliation{Institute of Physics, University of Silesia, 40-007 Katowice, Poland}
\affiliation{Silesian Center for Education and Interdisciplinary Research, University of Silesia, 41-500 Chorz{\'o}w, Poland}
\author{Jerzy {\L}uczka}
\email[Correspondence to J.{\L}.]{ (e-mail: jerzy.luczka@us.edu.pl)}
\affiliation{Institute of Physics, University of Silesia, 40-007 Katowice, Poland}
\affiliation{Silesian Center for Education and Interdisciplinary Research, University of Silesia, 41-500 Chorz{\'o}w, Poland}
\author{Peter H{\"a}nggi}
\affiliation{Institute of Physics, University of Augsburg, 86135 Augsburg, Germany}
\affiliation{Nanosystems Initiative Munich, Schellingstr, 4, D-80799 M{\"u}nchen, Germany}
\footnotetext{Correspondence to J.{\L}. (e-mail: jerzy.luczka@us.edu.pl)}

\begin{abstract}
We study far from equilibrium transport of a periodically driven inertial Brownian particle moving in a periodic potential. As detected recently for a SQUID ratchet dynamics (Spiechowicz J. \mbox{\& {\L}uczka J.} \textit{Phys. Rev. E} \textbf{91}, 062104  (2015)), the mean square deviation of the particle position from its average may involve three distinct intermediate, although extended diffusive regimes: initially as superdiffusion, followed by subdiffusion and finally, normal diffusion in the asymptotic long time limit. Even though these anomalies are transient effects,  their lifetime can be many, many orders of magnitude longer than the characteristic time scale of the setup and 
turns out to be extraordinarily sensitive to the system parameters like temperature or the potential asymmetry. In the paper we reveal mechanisms of diffusion anomalies related to  ergodicity of the system, symmetry breaking of the periodic potential and ultraslow relaxation of the particle velocity towards its steady state. Similar sequences of the diffusive behaviours could be detected in various systems including, among others, colloidal particles in random potentials, glass forming liquids and granular gases. 
\end{abstract}
\maketitle
\footnotetext{Correspondence to J.{\L}. (e-mail: jerzy.luczka@us.edu.pl)}
\section{Introduction}
Transport processes on the microscale can exhibit features rather different from those encountered in the macroscopic world. On such a scale they are strongly influenced by the presence of ubiquitous fluctuations and ambient noise. Rather than fighting the role of ambient  fluctuations may be put those to work for selective  manipulation of matter and diverse beneficial transport phenomena. Recent examples of such intriguing, noise-assisted phenomena include stochastic resonance \cite{gammaitoni1998}, enhancement of transport efficiency \cite{spiechowicz2014pre, spiechowicz2015njp, spiechowicz2016jstatmech, leon2013}, amplification of diffusion \cite{schreier1998,reimann2001,reguera2002, spiechowicz2015chaos}, hyperdiffusion \cite{siegle2010} or the vast field of Brownian motor transport \cite{hanggi2009,astumian2002}. The latter constitutes an archetype scheme  in which, even in the absence of an externally applied bias, directed motion may emerge by rectifying environmental noise via the mechanism of  breaking of spatiotemporal symmetries of the system. This working principle may be seen as a key for understanding processes ranging from intracellular transport \cite{bressloff2013}, cancer cell metastasis \cite{mahmud2009}, transport of ions through nanopores \cite{vlassiouk2007, karnik2007}, colloidal particles and cold atoms in optical lattices \cite{rousselet1994,kettner, matthias2003, renzoni2003, renzoni2005, denisov2014}, as well as vortices and Josephson phase in superconductors \cite{lee1999, villegas2003, silva2006, zapata1996, spiechowicz2014prb, weiss2000, sterck2005}. In view of the widespread applications of Brownian ratchets, directed motion controllability has become a focal point of research in nonequilibrium statistical physics which inspired a plethora of new microscale devices displaying unusual transport features \cite{hernandez2004, costache2010,serreli2007,drexler2013,grossert2016}.

On the other hand, the phenomenon of diffusion  is ever growing  in attracting interest since Einstein's and Smoluchowski's groundbreaking studies \cite{einstein1905,smoluchowski1906}. This stems from the fact of the universal character and ubiquitous presence of diffusion in both regimes of classical transport  and as well  for  quantum setups \cite{chaos2005,sokolov2005}. One century after those pioneering Einstein-Smoluchowski papers a new class of systems emerged that does not obey the common Gaussian law of large numbers  \cite{lutz2013}. Their behavior is dominated by large and rare fluctuations that are described by nonexponential decay laws, commonly referred to as \emph{L\'evy statistics}. Other characteristic features of these systems are \emph{ergodicity breaking}, that is, non-equivalence between  time averages and the corresponding ensemble averages \cite{jeon2011, burov2011, meroz2015} as well as \emph{aging}, i.e. a manifest dependence of physical observable on the time span between initialisation of the system and the start of the measurement \cite{bouchaud1992, burov2010}. Both ergodicity breaking and aging are in essence two sides of the same coin as they are intimately connected to \emph{non-stationarity or ultra-slow relaxation} \cite{sokolov2012,barkai2012}. These systems exhibit various forms of \emph{diffusion anomalies} \cite{metzler2000, sokolov2005, metzler2014,goychuk2014, zaburdaev2015} which were also demonstrated in numerous experiments \cite{metzler2014, zaburdaev2015}. This kind of dynamics may not survive until the asymptotic long time regime, nonetheless, lately even its transient nature has been predicted theoretically and observed experimentally \cite{bronstein2009, hanes2012t, hanes2012e}.

Nevertheless, despite many years of intense and beneficial research in ratchet physics, a more detailed analysis of diffusion in such systems has been addressed only recently \cite{spiechowicz2015pre}. There, an archetypal model of the Brownian ratchet was studied revealing that the mean square deviation of the particle position from its average proceeds within three intermediate stages: initially starting as superdiffusion, followed next as subdiffusion and finally as normal diffusion in the asymptotic long time limit. These transient diffusion anomalies may last many orders longer than the characteristic time scales of the setup and their lifetime can be controlled by variation of the system temperature and its asymmetry. However, the mechanism behind this phenomenon as well as its controllability remains somewhat moot, thus being far from fully understood. The detected sequence of superdiffusion-subdiffusion-normal diffusion is not an exception and seems to be universal for some  systems. A similar pattern has been observed also in other setups like Brownian particles in random potentials \cite{hanes2012t,hanes2012e,goychuk2014}, granular gases in a homogeneous cooling state \cite{bodrova2012} and glass forming liquids \cite{haxton2010}. While the above systems are significantly different, their diffusion behaviour is very similar. In this paper we base our study on a simple ratchet model to explain the underpinning mechanism responsible for such a sequence of diffusion anomalies. We analyze the extraordinary properties of systems exhibiting anomalous diffusion like ergodicity breaking and nonstationarity to establish an interrelation between directed ratchet transport and diffusion anomalies.
\section{Model}
\begin{figure}[t]
	\centering
	\includegraphics[width=0.4\linewidth]{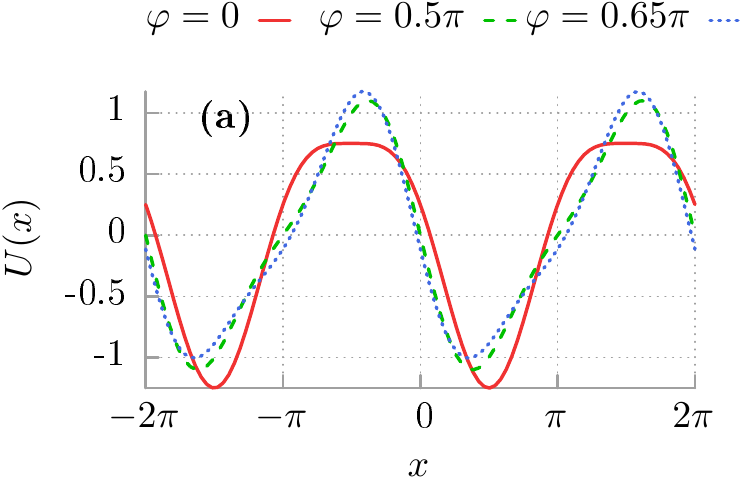}
	\includegraphics[width=0.4\linewidth]{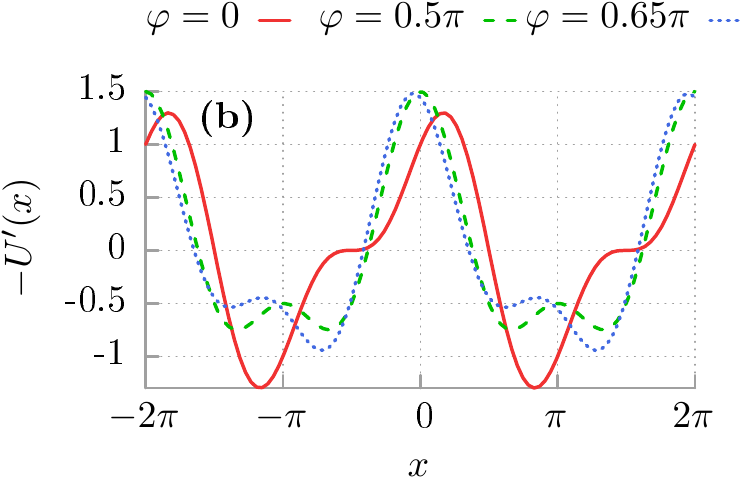}
	\caption{Panel (a): The potential Eq. (\ref{eq:potential}) for $\Delta U = 1$ and $L = 2\pi$ depicted in the symmetric case $\varphi = 0$ in comparison with ratchet one for two values of the potential asymmetry parameter $\varphi = 0.5\pi$ and $\varphi = 0.65\pi$. Panel (b): the conservative potential force $-U'(x)$.}
	\label{fig1}
\end{figure}

We consider the generic model of a ratchet system which consists of (i) a classical inertial particle of mass $M$, (ii) moving in a deterministic asymmetric ratchet potential $U(x)$, (iii) driven by an unbiased time-periodic force $A\cos{(\Omega t)}$ of amplitude $A$ and angular frequency $\Omega$, and (iv) subjected to thermal noise of temperature $T$. The corresponding Langevin equation reads \cite{spiechowicz2015pre}
\begin{equation}
	\label{eq:model}
	M\ddot{x} + \Gamma\dot{x} = -U'(x) + A\cos{(\Omega t)} + \sqrt{2\Gamma k_B T}\,\xi(t),
\end{equation}
where the dot and the prime denote differentiation with respect to  time $t$ and the Brownian particle coordinate $x$, respectively. The parameter $\Gamma$ stands for the friction coefficient and $k_B$ is the Boltzmann constant. Thermal equilibrium fluctuations are modeled by $\delta$-correlated, Gaussian white noise $\xi(t)$ of zero mean and unit intensity, i.e.,
\begin{equation}
	\langle \xi(t) \rangle = 0, \quad \langle \xi(t)\xi(s) \rangle = \delta(t-s).
\end{equation}
The spatially periodic potential $U(x)$ is assumed to be in a double-sine form \cite{jung1996} of period $2\pi L$ and a barrier height $\Delta U$, namely
\begin{equation}
	\label{eq:potential}
	U(x) = -\Delta U\left[ \sin{\left(\frac{x}{L}\right)} + \frac{1}{4}\sin{\left( 2 \frac{x}{L} + \varphi - \frac{\pi}{2}\right)}\right].
\end{equation}
It is \emph{reflection-symmetric} whenever there exists a shift $x_0$ such that $U(x_0 + x) = U(x_0 - x)$ for any $x$. The relative phase $\varphi$ between the two harmonics serves as a control parameter of the reflection-asymmetry of this potential. If $\varphi \neq 0$ then generally its reflection symmetry is broken which we in turn classify as as a \emph{ratchet}-device \cite{hanggi2009,astumian2002}, cf. Fig. \ref{fig1}. The studied model describes a wealth of physical systems, as cited in the introductory section.

As  only relations between scales of length, time and energy are relevant but not their absolute values we next formulate the above equations of motion in its dimensionless form. To do so, we first introduce the characteristic dimensionless scales for the  system under consideration \cite{machura2008}
\begin{equation}
	\hat{x} = \frac{x}{L}, \quad \hat{t} = \frac{t}{\tau_0}, \quad \tau_0 = \frac{\gamma L^2}{\Delta U},
\end{equation}
so that the dimensionless form of the Langevin dynamics (\ref{eq:model}) reads
\begin{equation}
	\label{eq:dimlessmodel}
	m\ddot{\hat{x}} + \dot{\hat{x}} = -\hat{U}'(\hat{x}) + a\cos{(\omega \hat{t})} + \sqrt{2Q} \hat{\xi}(\hat{t})\;.
\end{equation}
Here, the dimensionless potential $\hat{U}(\hat{x}) = U(x)/\Delta U = U(L\hat{x})/\Delta U = \hat{U}(\hat{x} + 2\pi)$ possesses the period $2\pi$ and the unit barrier height is $\Delta U = 1$. Other parameters are: $m = M/(\gamma\tau_0)$, $a = (L/\Delta U)A$, $\omega = \tau_0\Omega$. The rescaled thermal noise reads \mbox{$\hat{\xi}(\hat{t}) = (L/\Delta U)\xi(t) = (L/\Delta U)\xi(\tau_0\hat{t})$} and assumes  the same statistical properties, namely $\langle \hat{\xi}(\hat{t}) \rangle = 0$ and \mbox{$\langle \hat{\xi}(\hat{t})\hat{\xi}(\hat{s}) \rangle = \delta(\hat{t} - \hat{s})$}. The dimensionless noise intensity $Q = k_BT/\Delta U$ is the ratio of thermal and the activation energy the particle needs to overcome the nonrescaled potential barrier. From now on we will use only the dimensionless variables and shall omit the hat in all quantities appearing in the Langevin equation (\ref{eq:dimlessmodel}). We stress that all forces on the right hand side of Eq. (\ref{eq:dimlessmodel}) are non-biased: the spatial period-average of $U'(x)$, the temporal period average of $a \cos(\omega t)$ and the noise average of $\xi(t)$  are zero. 
\subsection{Quantifiers of diffusive transport}
There exist suitable quantifiers which characterize the Brownian particle diffusion and spread of its trajectories. The most common one  is the mean square deviation (or variance) of the  coordinate degree of freedom $x(t)$, namely,
\begin{equation}
	\label{Delta}
	\langle \Delta x^2(t) \rangle = \langle [ x(t) - \langle x (t) \rangle ]^2 \rangle = \langle x^2(t) \rangle - \langle x(t) \rangle^2,
\end{equation}
where the averaging is over all possible thermal noise realizations as well as over initial conditions for the position $x(0)$ and the velocity $\dot x(0)$. The latter is mandatory due to the fact that especially in the deterministic limit of vanishing thermal fluctuations intensity $Q \to 0$ the dynamics may not be ergodic and corresponding results may be affected by a specific choice of those initial conditions. Even though the diffusive  motion of $x(t)$ deviates at intermediate times  from being normal it is nevertheless appealing   to introduce a time-dependent "diffusion coefficient" $D(t)$, reading \cite{khoury2011,mei2014,spiechowicz2015pre}
\begin{equation}
	\label{D(t)}
	D(t) = \frac{\langle \Delta x^2(t) \rangle}{2t} \;.
\end{equation}
For the diffusion process, the asymptotic time evolution of the  mean square deviation (MSD) $\langle \Delta x^2(t) \rangle$ becomes an increasing function of elapsing time and typically grows according to a power law \cite{metzler2014, zaburdaev2015}
\begin{equation}
	\langle \Delta x^2(t) \rangle \sim t^{\alpha}.
\end{equation}
The exponent $\alpha$ specifies a type of anomalous diffusion. Normal diffusion characterized for  $\alpha = 1$. The two distinct regimes of anomalous diffusion are quantified  as \cite{metzler2014, zaburdaev2015}: subdiffusion if $0 < \alpha < 1$ and superdiffusion if $\alpha > 1$. In the former, subdiffusive case the MSD increases over time slower than normal while for superdiffusion it is growing faster than  normal diffusion. The time dependent diffusion coefficient $D(t)$ therefore allows to differentiate  between these anomalous intermediate regimes of diffusion; namely superdiffusion occurs when $D(t)$ increases, the case of decreasing $D(t)$ corresponds to subdiffusion and for $D(t)= const.$ normal diffusion occurs. Only when asymptotically $\alpha$ approaches unity the time-independent diffusion coefficient $D$ is given by
\begin{equation}
	D = \lim_{t \to \infty} D(t).
\end{equation}
To gain insight into the origin of the diverse diffusion phenomena in the system under consideration it helps to study the occurrence  of non-vanishing directed transport, that is the noise and time averaged  velocity $\mathbf{v}(t)$, defined by the relation
\begin{equation}
	\label{v(t)}
	\mathbf{v}(t) =  \frac{1}{\mathsf{T}} \int_{t}^{t + \mathsf{T}} ds\, \langle \dot{x}(s)\rangle\;, 
\end{equation}
where  $\mathsf{T}=2\pi/\omega$ is a period of  the time-periodic force. 
In the asymptotic long time limit this quantity becomes time-independent,  while the noise-averaged quantity alone assumes a time-periodic function of the asymptotic time-periodic phase-space probability. Put differently, in the asymptotic long time limit, the mean velocity $\langle \dot{x}(t)\rangle$ takes the form of a Fourier series over all possible higher harmonics \cite{gammaitoni1998,jung1993}; yielding
\begin{equation}
	\lim_{t \to \infty} \langle \dot{x}(t) \rangle = \mathbf{v} + v_{\omega}(t) + v_{2\omega}(t) + ...
\end{equation}
where $\mathbf{v}$ is the  time-independent (dc) component while $v_{n\omega}(t)$ denote time-periodic higher harmonic functions of zero average over the  fundamental period $\mathsf{T}=2\pi/\omega$ of the driving. The dc component $\mathbf{v}$ is  obtained as
\begin{equation}
	\label{v}
	\mathbf{v} = \lim_{t \to \infty} \mathbf{v}(t).
\end{equation}
Due to the presence of the external driving the Brownian particle is taken far away from thermal equilibrium and a time-dependent nonequilibrium state is reached in the asymptotic long time regime. Since all forces in the right hand side of Eq. (\ref{eq:dimlessmodel}) are non-biased, a necessary condition for the occurrence of directed transport $\mathbf{v} \neq 0$ is the breaking of the reflection symmetry of the potential $U(x)$ \cite{hanggi2009,astumian2002}.
\begin{figure}[t]
	\centering
	\includegraphics[width=0.4\linewidth]{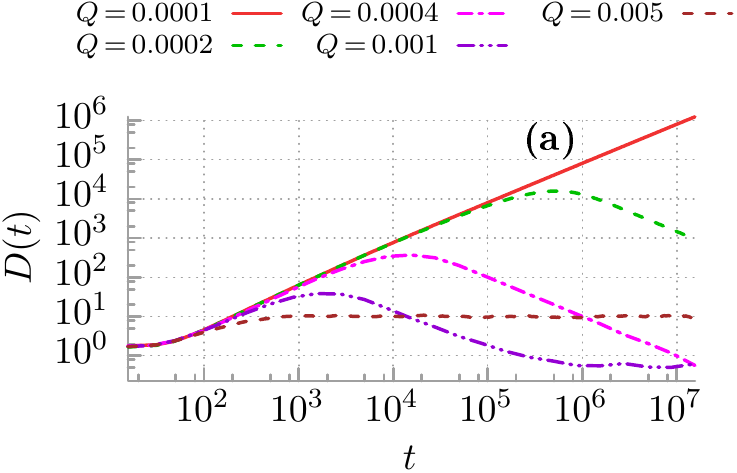}
	\includegraphics[width=0.4\linewidth]{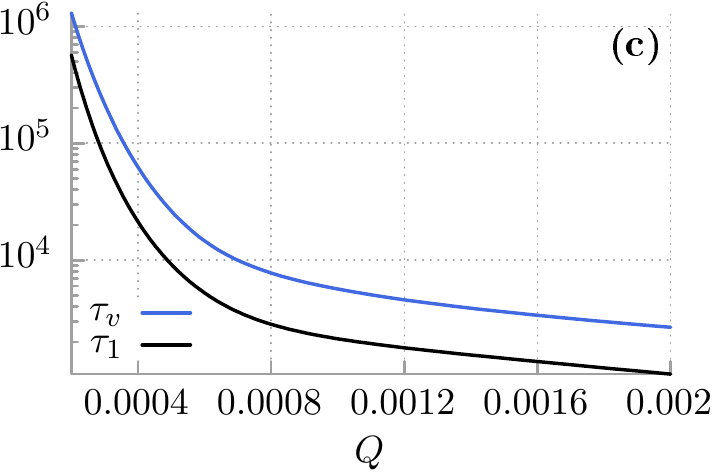}\\
	\includegraphics[width=0.4\linewidth]{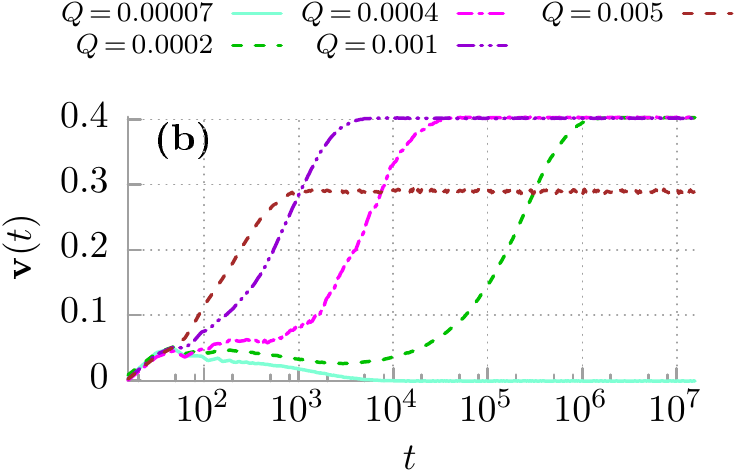}
	\includegraphics[width=0.4\linewidth]{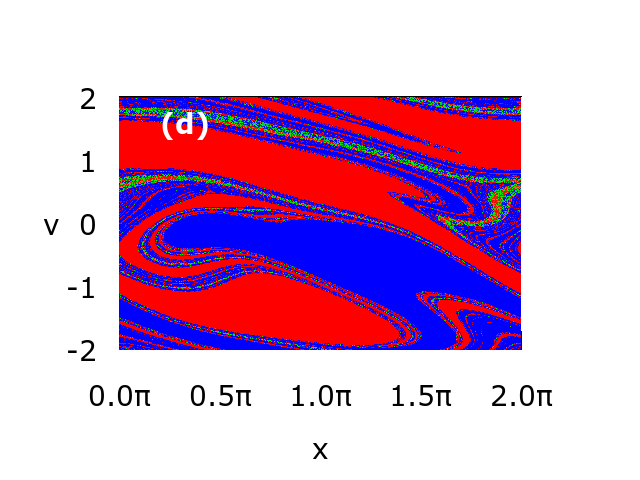}
	\caption{Control of anomalous diffusion regimes by temperature 	$Q \propto T$ in the ratchet potential Eq. (3) with $\varphi =\pi/2$. Panel (a): the diffusion coefficient $D(t)$. Panel (b): relaxation of the velocity $\mathbf{v}(t)$ to its nonequilibrium stationary state. Panel (c): the crossover time $\tau_1$ separating superdiffusion and subdiffusion stages of the diffusion process and the velocity relaxation time $\tau_v$ versus system temperature. Panel (d): basins of attraction for the asymptotic long time particle velocity $\mathbf{v}$. Red colour indicates the running states with the positive velocity $\mathbf{v} = 0.4$, blue corresponds to its negative counterpart $\mathbf{v} = -0.4$ and green marks the locked states $\mathbf{v} \approx 0$. Parameters are: $m = 6$, $a = 1.899$, $\omega = 0.403$ and $Q = 0.0004$.}
	\label{fig2}
\end{figure}
\section{Anomalous diffusion: The role of finite temperature}
Even with the help of our innovative computational simulations, see the section Methods, the system described by Eq. (\ref{eq:dimlessmodel}) is too complex to analyze numerically in a systematic manner. It possess a five-dimensional parameter space $\{m, a, \omega, \varphi, Q\}$. However, in our recent paper \cite{spiechowicz2015pre} we revealed the remarkable regime of thermal noise induced ratchet effect, (cf. Fig. 6 in Ref.  \cite{spiechowicz2015njp}): At zero temperature $Q=0$ the averaged velocity $\mathbf{v} \approx 0$. If $Q$  starts to increase, the velocity $\mathbf{v}$ also increases attaining  the maximal value $\mathbf{v} \approx 0.4$ for $Q \approx 0.0004$ and next it decreases to zero as $Q$ grows. It is so for $\varphi =\pi/2$ in the potential (3) implying its most asymmetrical ratchet form. In this regime, evolution of the mean square deviation can be divided into three time domains as depicted in Fig. \ref{fig2}(a) (see also Fig. 3 in \cite{spiechowicz2015pre}): the early period of superdiffusion $\tau_1$ (where $D(t)$ is an increasing function of time), the intermediate interval $\tau_2$ where subdiffusion is developed (there $D(t)$ is a decreasing function of time) and approaches the asymptotic long time regime where normal diffusion occurs (i.e. $D(t) \approx$ constant). In Fig. \ref{fig2}(c) we show how the crossover time $\tau_1$ depends on temperature. The lifetime $\tau_1$ of the superdiffusion regime is extremely long in the low temperature limit, c.f. the curve corresponding to $Q = 0.0001$ in Fig. 2(a). However, persistent superdiffusion occurs only in the deterministic case when formally $Q = 0$. The deflection of this stage of diffusion can expressively be noted as temperature increases. For sufficiently high temperature the motion is initially superdiffusive and next normal diffusion occurs, see the case of $Q = 0.005$ in Fig. \ref{fig2}(a).
\subsection{Strong ergodicity breaking}
To explain the above diffusion anomalies let us now study the deterministic dynamics $Q = 0$ and the corresponding structure of basins of attraction for the asymptotic long time velocity $\mathbf{v}$. The result is shown in Fig. \ref{fig2}(d). There exist only three attractors: the set $\mathbf U_+$ which consists of all running states with positive velocity $\mathbf{v} \approx  0.4$ (marked by red colour), the set $\mathbf U_-$ of states running with negative velocity $\mathbf{v} \approx -0.4$ (marked by blue colour) and the set $\mathbf U_0$ of locked states $\mathbf{v} \approx 0$ when the motion is bounded to a finite number of the potential wells (green colour).
This simple structure is crucial for the occurrence of initial superdiffusive (ballistic) stage of motion \cite{jung1996}. Roughly speaking, there are three classes of trajectories: $x(t)\sim 0.4 t$, $x(t) \sim -0.4 t$ and $x(t) \sim 0$. Moreover, the running states of either positive or negative velocity occupy almost even area of the initial phase space of the system. This fact causes that the contribution of the averaged trajectory to the mean-square deviation $\langle \Delta x^2(t) \rangle   = \langle x^2(t) \rangle - \langle x(t) \rangle^2$ is negligible small $\langle x(t) \rangle \approx 0$ in comparison to its second moment $\langle x^2(t) \rangle \sim t^2$. As a consequence $\langle \Delta x^2(t) \rangle \sim t^2$ also for the asymptotic long time limit $t\to \infty$. Therefore then the superdiffusive regime is persistent and its lifetime $\tau_1 \to \infty$ when $Q \to 0$. In fact, it is an example of \emph{ergodicity breaking} \cite{meroz2015}. The phase space of the system consists of three ($v \pm 0.4$, $v \approx 0$) disjoint sets which are mutually inaccessible. In other words a phase point starting e.g. in the blue region in Fig. \ref{fig2}(d) and evolving according to Eq. (\ref{eq:dimlessmodel}) will stay in this region forever and cannot penetrate the other two sets. As a consequence, statistical properties of the system deduced from time average along one sufficiently long trajectory are not the same as determined by the ensemble averaging. Such a decomposability of the phase space is called \emph{strong ergodicity breaking} \cite{meroz2015} in the language accepted in the community of nonequilibrium statistical physicists. This is essentially a refinement of original Boltzmann's definition and in other words means that a trajectory associated with time evolution of an initial condition cannot pass arbitrarily close to any point of a phase-space, put simply initial conditions are never fully forgotten - a property which finds its analogue in any definition of ergodicity.
\begin{figure}[t]
	\centering
	\includegraphics[width=0.32\linewidth]{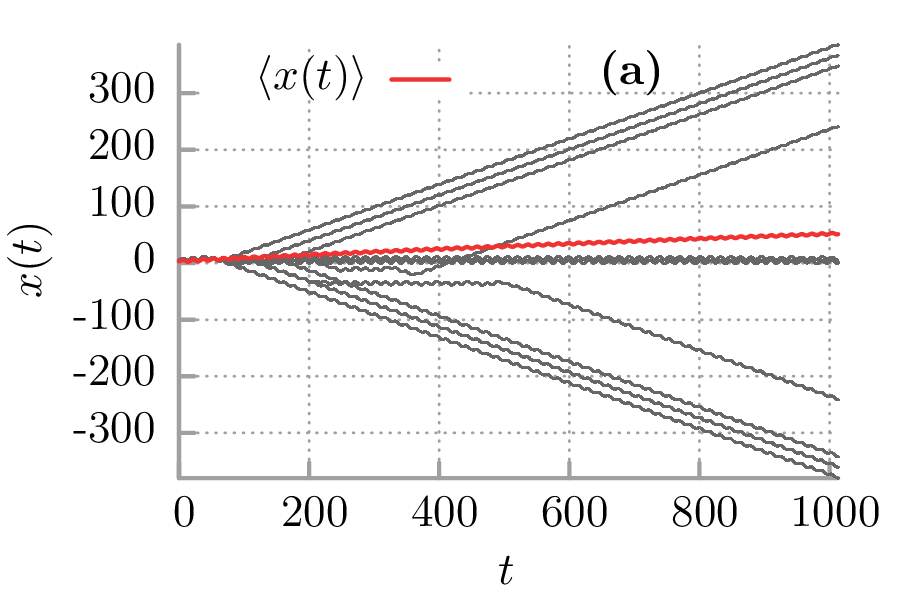}
	\includegraphics[width=0.32\linewidth]{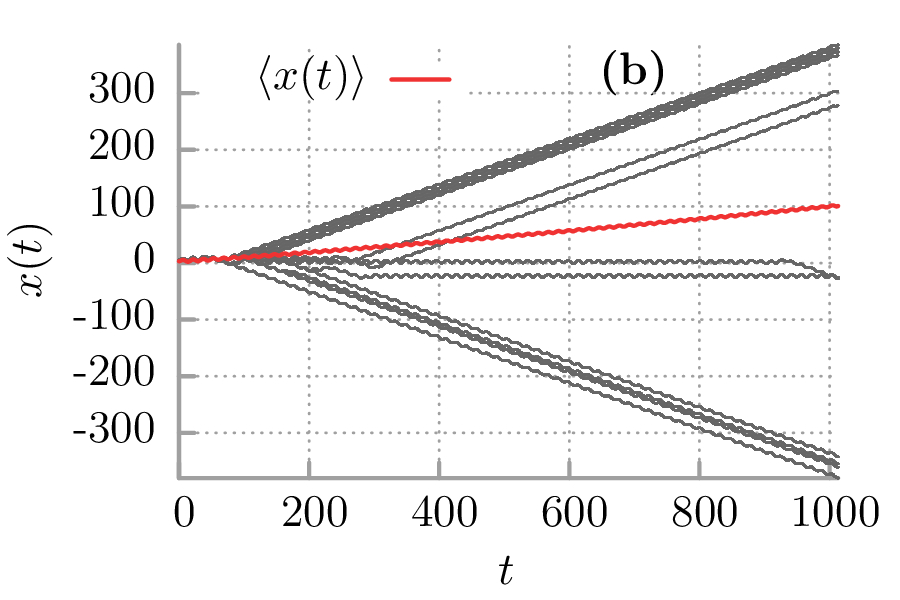}
	\includegraphics[width=0.32\linewidth]{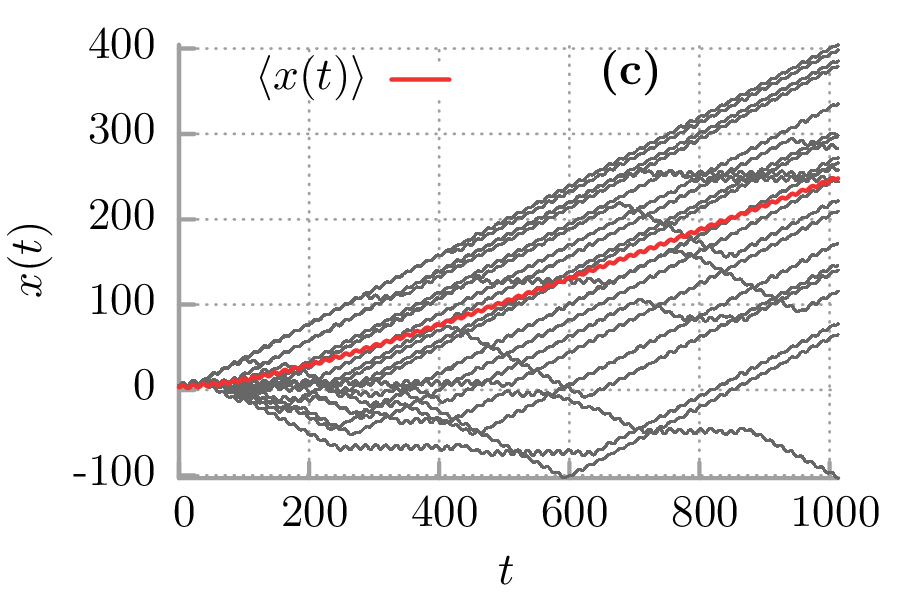}
	\caption{Various typical sample trajectories of the Brownian particle dynamics. These are depicted for different temperature of the system $Q \propto T$. Panel (a), (b) and (c) corresponds to $Q = 0.0001$, $Q = 0.0004$ and $Q = 0.005$, respectively. Other parameters are the same as in Fig. \ref{fig2}.}
	\label{fig3}
\end{figure}
\begin{figure}[t]
	\centering
	\includegraphics[width=0.32\linewidth]{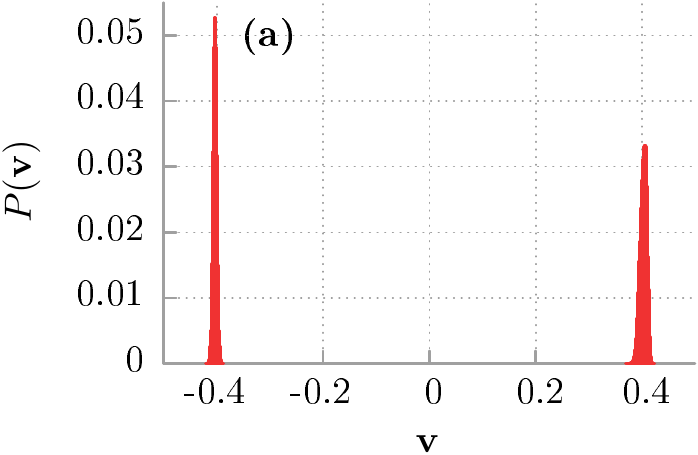}
	\includegraphics[width=0.32\linewidth]{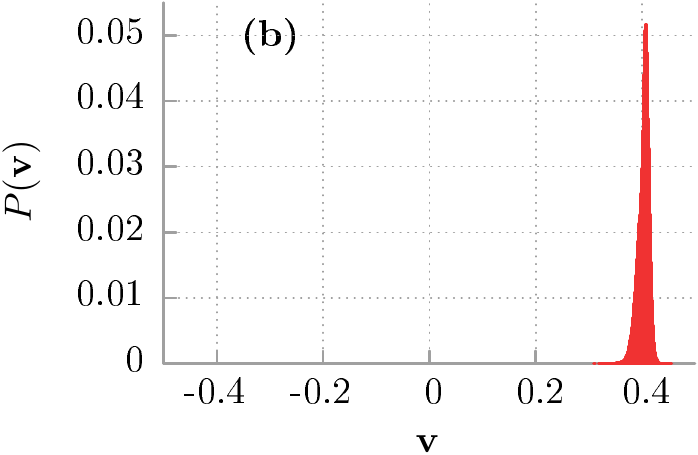}
	\includegraphics[width=0.32\linewidth]{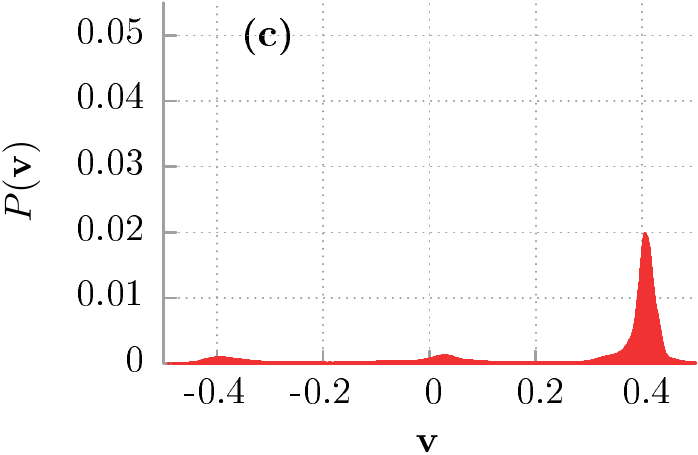}
	\caption{The probability distribution $P(\mathbf{v})$ of the asymptotic long time particle velocity $\mathbf{v}$ are presented for different values of temperature  $Q \propto T$. Panel (a): $Q = 0.0001$, (b): $Q = 0.0004$ and (c): $Q = 0.005$. Other parameters are the same as in Fig. \ref{fig2}.}
	\label{fig4}
\end{figure}
\subsection{Superdiffusion versus thermal fluctuations}
We expect that thermal noise with the intensity $Q\neq 0$ perturbs  deterministic dynamics and equilibrium fluctuations enable stochastic escape events connecting coexisting deterministic disjoint attractors \cite{hanggi1990}. Therefore in a presence of  noise the system is  ergodic \cite{spiechowicz2015pre}. In \mbox{Fig. \ref{fig3}} we depict 1024 sample Brownian particle trajectories for different thermal fluctuations intensity $Q \propto T$. Let us consider for definiteness an initial condition $(x_0, v_0)$ for which in the deterministic case $Q = 0$ the particle moves along the trajectory $x(t) \sim -0.4 t$. When $Q > 0$, from time to time the particle is kicked by thermal fluctuations and escapes from its initial trajectory jumping on the other like $x(t) \sim 0.4 t$ or $x(t) \sim 0$. After some time, the particle can again break away the ongoing trajectory and jump on the different one. This process reduces a number of trajectories which contribute to superdiffusion (here $x(t) \sim -0.4 t$) and destroys the initial simple structure of only three attractors. It can be directly seen in Fig. \ref{fig4} where we present the probability distributions $P(\mathbf{v})$ of the asymptotic long time  velocity $\mathbf{v}$ for different values of temperature $Q \propto T$. In the deterministic limit of vanishing thermal noise intensity $Q = 0$ there are two almost equally pronounced $\delta$-peaks representing the running solutions of either positive $\mathbf{v} = 0.4$ or negative $\mathbf{v} = -0.4$ direction. The deviation of this structure of states is distinctly noticeable as temperature increases. Thermal noise first blurs $\delta$-peaks observed in the deterministic case and then eliminates both the negative running as well as locked states in favor of the positive ones $\mathbf{v} \approx 0.4$. Further increase in temperature causes additional smearing of the probability distribution $P(\mathbf{v})$. Clearly, in presence of thermal noise the system forgets about its initial conditions so that ergodicity is restored.

Overall, if temperature grows the jumps between different types of solutions are more and more frequent (as in \mbox{Fig. \ref{fig3}(c)}) and the mean time to destroy the deterministic structure of attractors becomes shorter. As a consequence, the lifetime $\tau_1$ of superdiffusion is also shortened. It is now obvious that $\tau_1$ is a decreasing function of temperature $Q \propto T$ in accordance with results presented in Fig. \ref{fig2}(a) and Fig. \ref{fig2}(c).
\subsection{Superdiffusion versus velocity relaxation}
Now, in order to quantify the mean time to destroy the deterministic structure of attractors we study a relaxation process of the velocity degree of freedom. The result is presented in Fig. \ref{fig2}(b) where time evolution of the velocity $\mathbf{v}(t)$ defined in Eq. (\ref{v(t)}) is depicted. In the limiting regime of weak thermal noise     the velocity quickly relaxes to zero $\mathbf{v}(t)\approx 0$, see e.g. the case $Q = 7 \cdot 10^{-5}$. It means that at least during the data acquisition time in the numerical experiment $\mathcal{T} \approx 10^7$ equilibrium fluctuations are not able to wipe out the deterministic structure of basins of attraction so that the system still remembers three solutions and the mean velocity is zero.  However, since in a presence of thermal noise the attractors coexisting in the phase space are connected via stochastic escape events we expect that even if temperature is low a relevant trajectory will eventually fully sample the available state space of the system. For moderate and strong thermal fluctuations the system forgets about its initial conditions since the velocity relaxes to the nonzero steady state $\mathbf{v} = 0.4$, c.f. the case $Q=0.005$ in Fig. \ref{fig2}(b). Its time significantly influences the duration of superdiffusion $\tau_1$ as we demonstrate in Fig. \ref{fig2}(c). A rough synchronization of these two curves is observed there. It is remarkable that for the considered  case of $\varphi = 0.5\pi$ these two characteristic times increase over three orders of magnitude when the intensity $Q$ of thermal fluctuations varies in the small interval $[0.0002,0.01]$. The bottom bound of this window is chosen so that thermal noise is still able to destroy the deterministic structure of coexisting regular attractors in the data acquisition time $\mathcal{T} \approx 10^7$.
\begin{figure}[t]
	\centering
	\includegraphics[width=0.4\linewidth]{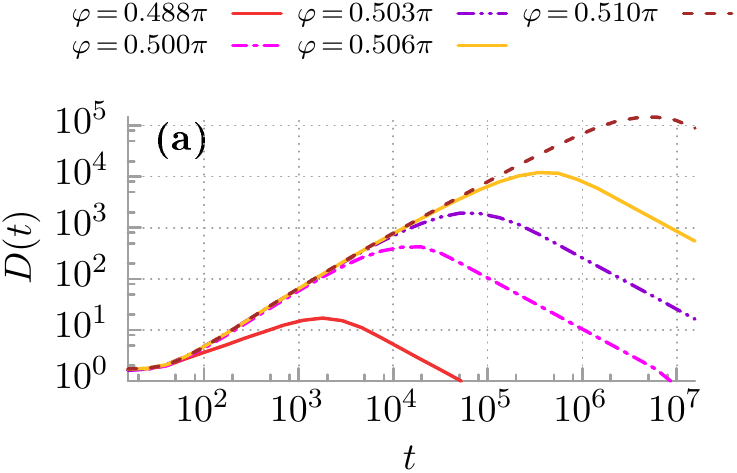}
	\includegraphics[width=0.4\linewidth]{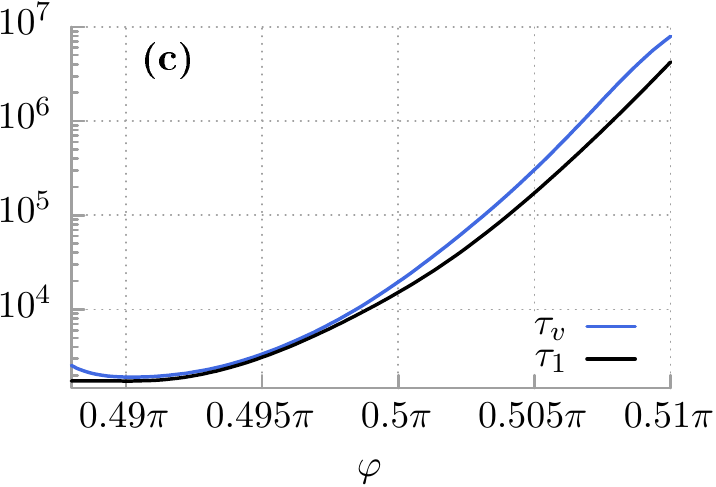}\\
	\includegraphics[width=0.4\linewidth]{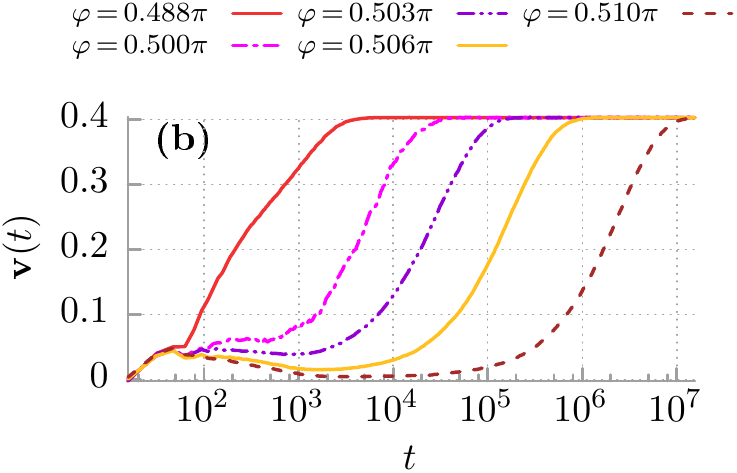}
	\includegraphics[width=0.4\linewidth]{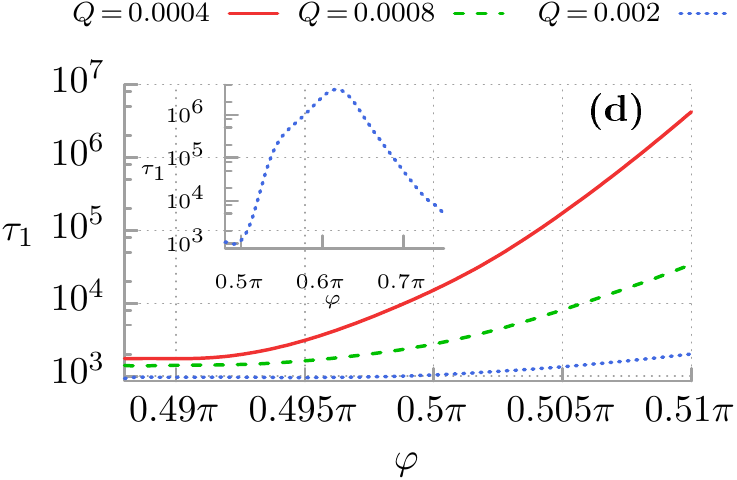}
	\caption{Impact of the asymmetry parameter $\varphi$ of the potential $U(x)$ on the diffusion process. Panel (a): sensitivity of the diffusion coefficient $D(t)$ in the vicinity of $\varphi = \pi/2$. Panel (b): relaxation of the velocity $\mathbf{v}(t)$ to its nonequilibrium steady state. Panel (c): the crossover time $\tau_1$ separating superdiffusion and subdiffusion stages of the diffusion process and the velocity relaxation time $\tau_v$ versus potential asymmetry parameter $\varphi$. Panel (d): the crossover time $\tau_1$ versus $\varphi$ depicted for different values of temperature $Q \propto T$. In the inset $\tau_1$ is shown for $Q = 0.002$ in larger interval of $\varphi$ capturing its maximum. Other parameters are the same as in Fig. \ref{fig2}.}
	\label{fig5}
\end{figure}
\section{Anomalous diffusion: Impact of potential asymmetry}
The sequence of superdiffusion-subdiffusion-normal diffusion is detected only when reflection symmetry of the potential $U(x)$ is broken, i.e. when it has a ratchet form. We already demonstrated that the initial superdiffusive stage of motion stems from the deterministic structure of the coexisting counter-propagating regular directed transporting attractors, see also the discussion on the power exponent $\alpha$ in \cite{spiechowicz2015pre}. Fig. \ref{fig5}(a) presents the impact of the potential asymmetry parameter $\varphi$ on the diffusion coefficient $D(t)$ in the vicinity of $\varphi = \pi/2$ where the crossover times $\tau_1$ and $\tau_2$ are extremely sensitive to its changes. Panel (b) of the same figure illustrates time evolution of the velocity $\mathbf{v}(t)$ for selected values of $\varphi$. Again, astonishingly, this quantity relaxes to its nonequilibrium steady state with the relaxation time $\tau_v$ strongly dependent on the magnitude of the potential asymmetry $\varphi$. Similarly as in the case of temperature influence, one can detect an interrelation between the superdiffusion crossover time $\tau_1$ and the velocity relaxation time $\tau_v$ so that if the potential asymmetry parameter $\varphi$ increases both $\tau_1$ and $\tau_v$ grow in the synchronized way. This feature is better visualized in the panel (c) where we contrast the relaxation time $\tau_v$ together with the superdiffusion lifetime $\tau_1$ for the fixed temperature $Q = 0.0004$. Both of them vary in the similar way and are extraordinarily sensitive to alteration of the potential asymmetry $\varphi$: Small changes of the order $10^{-2}$ in $\varphi$ are accompanied by the giant increase of $\tau_1$ and $\tau_v$ of the order $10^4$.
\begin{figure}[t]
	\centering
	\includegraphics[width=0.4\linewidth]{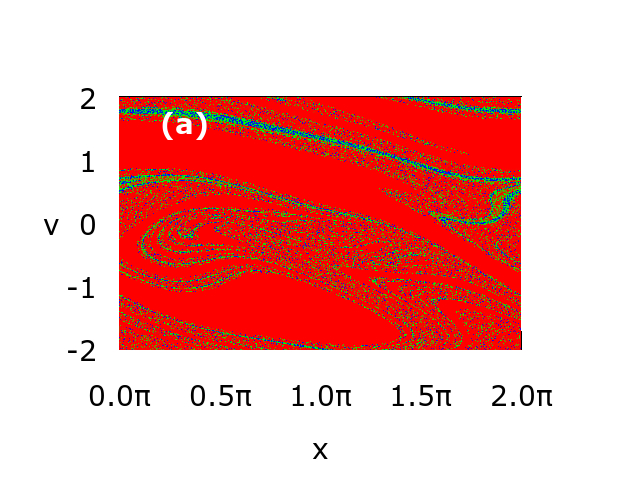}
	\includegraphics[width=0.4\linewidth]{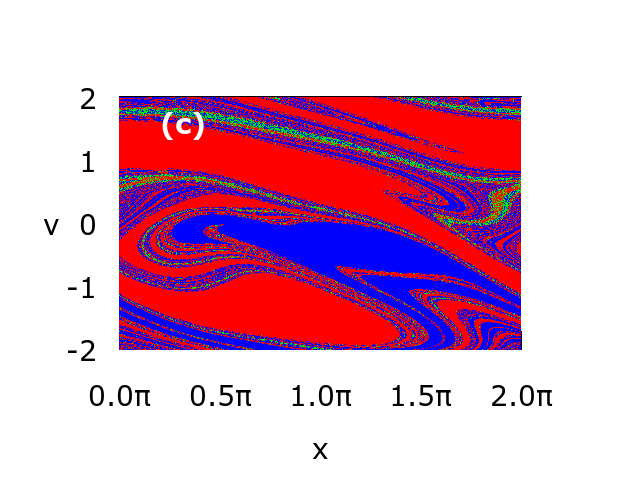} \\
	\includegraphics[width=0.4\linewidth]{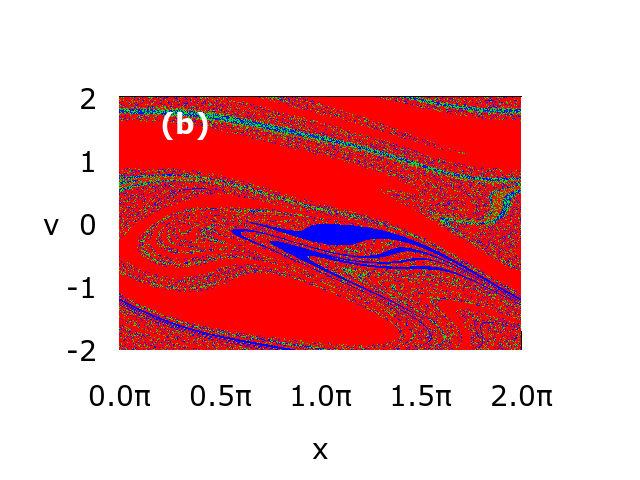}
	\includegraphics[width=0.4\linewidth]{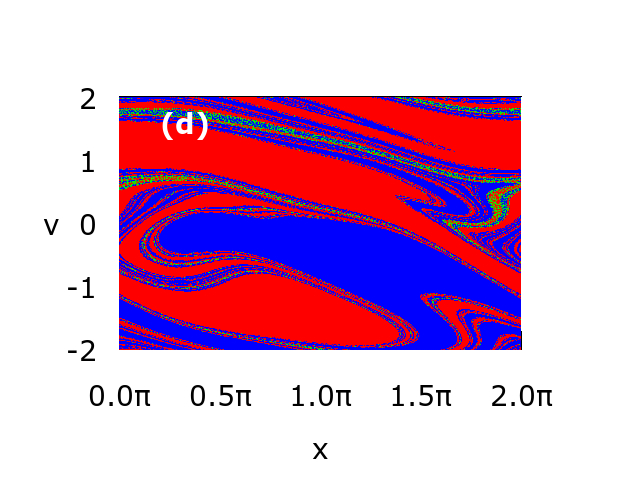}
	\caption{Basins of attraction for the asymptotic long time particle velocity $\mathbf{v}$ are presented for different values of the potential asymmetry parameter $\varphi$. Panel (a): $\varphi = 0.488\pi$, (b): $\varphi = 0.490\pi$, $\varphi = 0.494\pi$ and (d): $\varphi = 0.51\pi$. Red color indicates the running states with the positive velocity $\mathbf{v} = 0.4$, blue corresponds to its negative counterpart $\mathbf{v} = -0.4$ and  green color marks the locked states $\mathbf{v} \approx 0$. Other parameters are the same as in Fig. \ref{fig2}, except thermal noise intensity which is fixed to zero $Q = 0$.}
	\label{fig6}
\end{figure}
\subsection{Weak ergodicity breaking}
The time evolution of the diffusion coefficient $D(t)$ illustrated in Fig. \ref{fig2}(a) and Fig. \ref{fig5}(a) is very similar. However, if temperature tends to zero $Q \propto T \to 0$ the superdiffusion crossover time monotonically increases to infinity $\tau_1 \to \infty$ and ergodicity is broken. On the other hand, when the asymmetry parameter $\varphi$ increases the crossover time $\tau_1$ monotonically grows reaching the maximal value and next it starts to decrease, see the inset in Fig. \ref{fig5}(d). In other words when $Q \neq 0$ the lifetime $\tau_1$ does not tend to infinity for any value of the potential asymmetry parameter $\varphi$. It means that ergodicity cannot be \emph{strictly} broken by change of $\varphi$ although $\tau_1$ may be extremely long. In such a case the whole phase space is  still accessible because of the presence of thermal fluctuations, however, the time after it is fully sampled might be enormously long. From an experimental point of view the problem of ergodicity is a matter of time scale. In practice, there is not much difference whether the system is nonergodic or ergodic but exhibiting an extremely slow relaxation. If this effect is pronounced one often speaks about \emph{weak ergodicity breaking} \cite{meroz2015}. This kind of behaviour can be quantified by the Deborah number $De$ \cite{reiner1964}
\begin{equation}
	De = \frac{\tau}{\mathcal{T}},
\end{equation}
which is a ratio of a relaxation time $\tau$ of a given observable and the time of observation $\mathcal{T}$. If it diverges $De \to \infty$ then the system gives the impression of being "frozen" and the question whether it may eventually sample the full state space becomes irrelevant. This can happen not only because $\mathcal{T}$ is short but also because $\tau$ is extremely long. In our case the system behaves as weakly nonergodic when the superdiffusion lifetime $\tau_1$ or the velocity relaxation time $\tau_v$ is sufficiently large so that the condition $De = \tau/\mathcal{T} \gg 1$ is satisfied.
\subsection{Superdiffusion versus potential asymmetry}
\begin{figure}[t]
	\centering
	\includegraphics[width=0.32\linewidth]{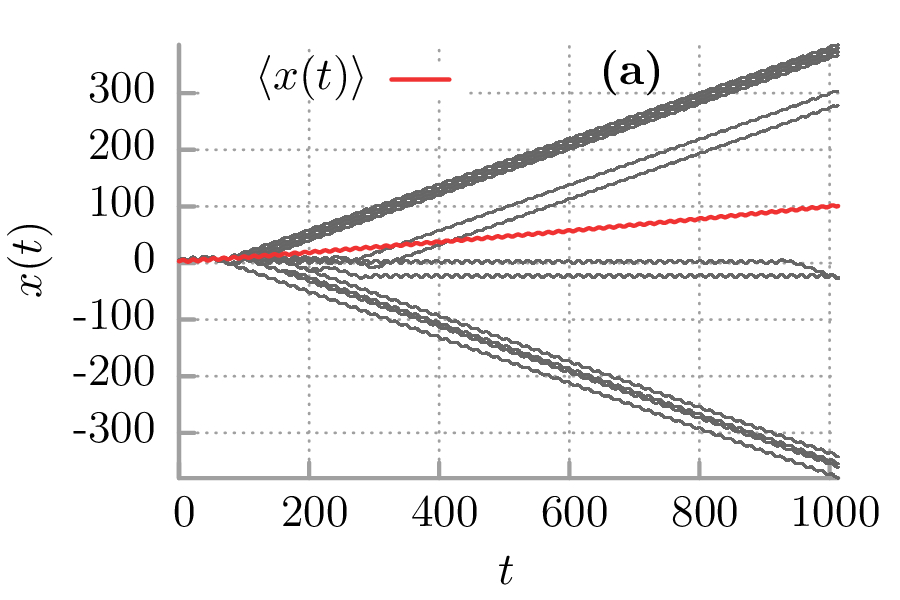}
	\includegraphics[width=0.32\linewidth]{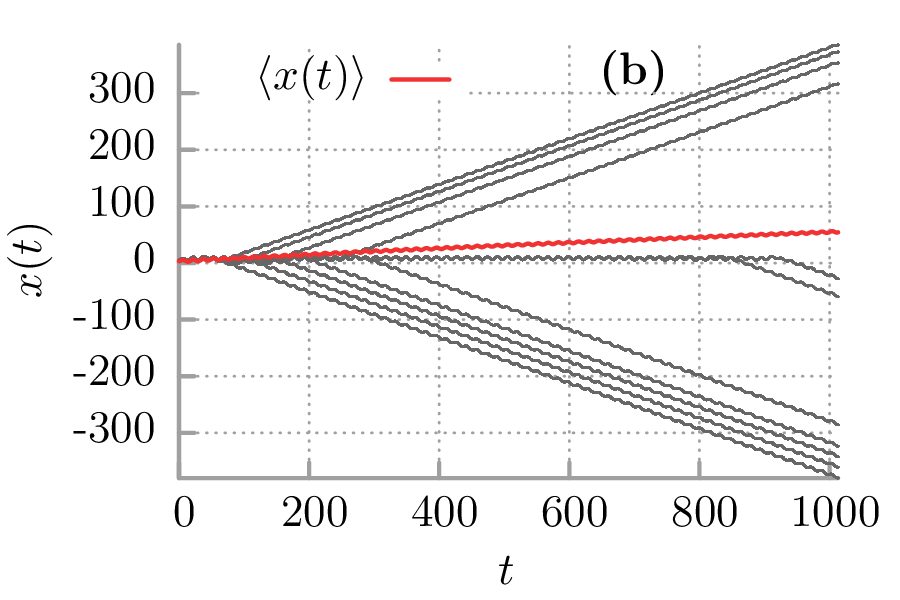}
	\includegraphics[width=0.32\linewidth]{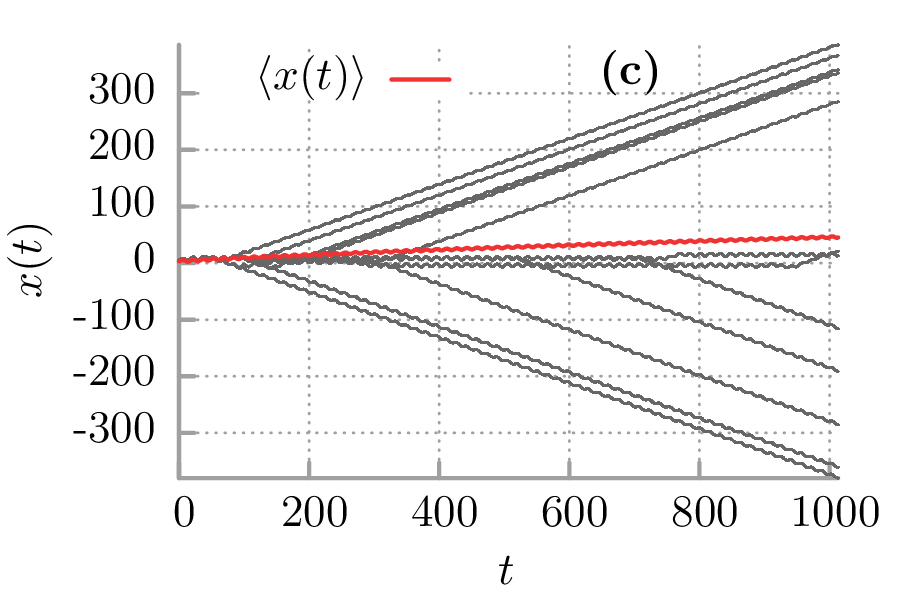}
	\caption{Characteristic sample trajectories of the Brownian particle depicted for different potential asymmetry parameters $\varphi$. Panel (a), (b) and (c) corresponds to $\varphi = 0.5\pi$, $\varphi = 0.505\pi$ and $\varphi = 0.51\pi$, respectively. Other parameters are the same as in Fig. \ref{fig2} and $Q = 0.0004$.}
	\label{fig7}
\end{figure}

\begin{figure}[t]
	\centering
	\includegraphics[width=0.32\linewidth]{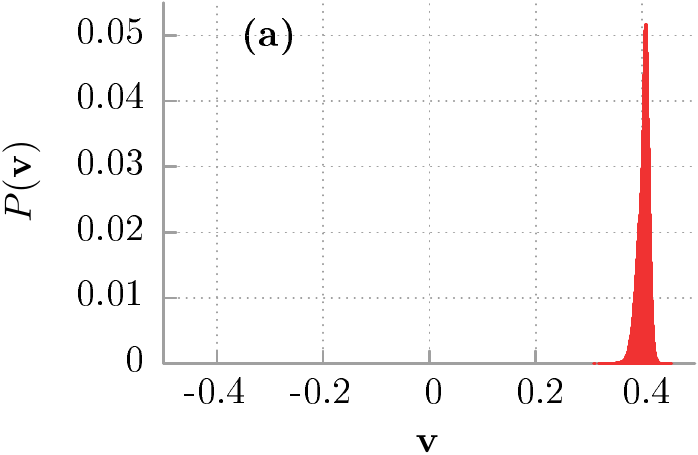}
	\includegraphics[width=0.32\linewidth]{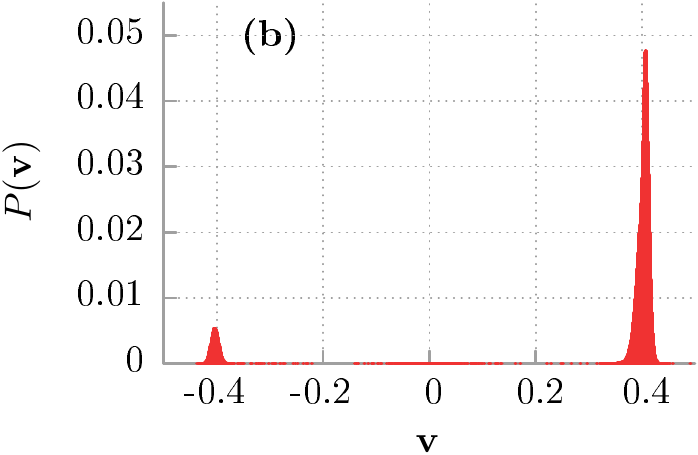}
	\includegraphics[width=0.32\linewidth]{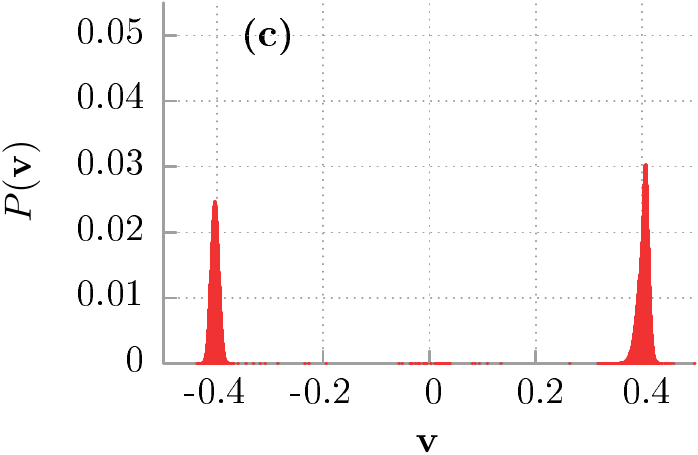}
	\caption{Probability distribution $P(\mathbf{v})$ of the asymptotic long time particle velocity $\mathbf{v}$ are presented for different values of the potential asymmetry parameter $\varphi$. Panel (a): $\varphi = 0.5\pi$, (b): $\varphi = 0.505\pi$ and (c): $\varphi = 0.51\pi$. Other parameters are the same as in Fig. \ref{fig2} and $Q = 0.0004$.}
	\label{fig8}
\end{figure}

To gain insight into the origin of the extremal sensitivity of the superdiffusion lifetime $\tau_1$ to changes of the potential asymmetry parameter $\varphi$ in the vicinity of $\varphi = \pi/2$ we now study variation of the structure of basins of attraction with respect to $\varphi$. In Fig. \ref{fig6} we present them in the deterministic limit $Q = 0$ and for different values of the parameter $\varphi$. Again, there are three classes of trajectories $\mathbf{v} = 0.4$, $\mathbf{v} = -0.4$ and $\mathbf{v} \approx 0$ with three corresponding basins of attraction $\mathbf{U}_+$, $\mathbf{U_-}$, $\mathbf{U}_0$ marked by red, blue and green colour, respectively. For $\varphi = 0.488\pi$ the set $\mathbf{U}_+$ is much larger than $\mathbf{U}_-$ and in consequence the superdiffusion lifetime $\tau_1$ is short. Surprisingly, we note that a tiny increase of the order $10^{-2}$ in the asymmetry parameter $\varphi$ is accompanied by a rapid growth of the set $\mathbf{U}_-$ of the initials conditions for which the asymptotic long time velocity $\mathbf{v}$ is negative $\mathbf{v} = -0.4$. Then the size of the sets $\mathbf{U}_+$ and $\mathbf{U}_-$ becomes comparable, there are more trajectories with the negative velocity and the spread of trajectories is large. Additionally, the noisy system needs more time  to  destroy this evident structure of basins of attraction via the mechanism of kicking out the particle from trajectories with the negative velocity $\mathbf{v} \approx -0.4$ onto those with positive ones $\mathbf{v} \approx 0.4$. Certainly, it is one of the reasons why the superdiffusion lifetime $\tau_1$ increases. 
However, we observe in the inset of Fig. \ref{fig5}(d) that the superdiffusion crossover time $\tau_1$ can still grow although the structure of basins of attraction for the asymptotic velocity $\mathbf{v}$ is not changing significantly any more, c.f. Fig. \ref{fig2}(d) and Fig. \ref{fig6}(d). We note that for $\varphi = \pi/2$,  asymmetry of the potential is maximal, the depth of the potential wells is maximal, mean first passage time $T_-$ to overcome the potential barrier to the left is much longer that mean first passage times $T_+$ to overcome the potential barrier to the right (not depicted). Further increase of $\varphi$ makes the potential less asymmetric,  its depth decreases, both $T_-$ and $T_+$ decrease and become comparable for $\varphi \approx 0.6$.  
Therefore its easier for thermal fluctuations to kick the particle onto the trajectory running in negative direction with the asymptotic velocity $\mathbf{v} <0$. We support this point of view by resource to simulation of sample trajectories for different values of the asymmetry parameter $\varphi$. The result is shown in Fig. \ref{fig7}. Indeed, we observe there that as $\varphi$ increases (i) progressively more particles move with the negative velocity and (ii) they start doing it earlier. These features are also visualized in Fig. \ref{fig8} where we present the probability distributions $P(\mathbf{v})$ of the asymptotic long time particle velocity  for different values of the potential asymmetry parameter $\varphi$. For fixed temperature $Q \neq 0$ an increase in $\varphi$ causes elimination of the positive running states in favour of the negative ones $\mathbf{v} \approx -0.4$. In turn, this mechanism is reversed with respect to that which occurred in the case of temperature changes, cf. Fig. 4.
\section{Role of initial velocity distribution}
\begin{figure}[t]
	\centering
	\includegraphics[width=0.32\linewidth]{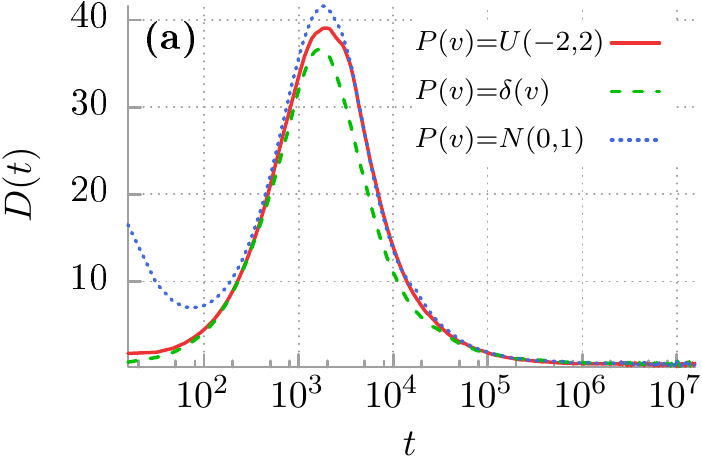}
	\includegraphics[width=0.32\linewidth]{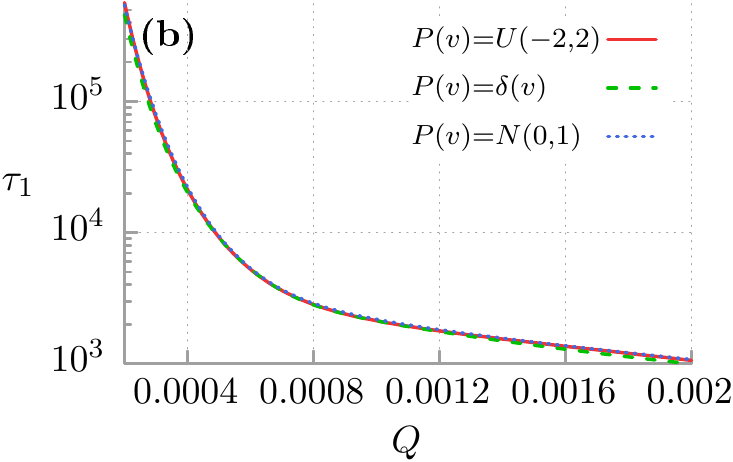}
	\includegraphics[width=0.32\linewidth]{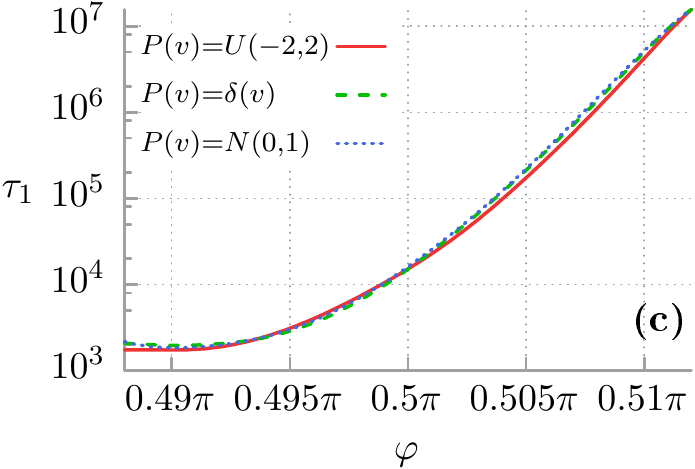}
	\caption{Impact of the initial probability distribution $P(v)$ for the Brownian particle velocity  on the diffusion anomalies observed in the system. Panel (a): the time dependent diffusion coefficient $D(t)$ for temperature $Q = 0.001$. Panel (b) and (c): the crossover time $\tau_1$ separating superdiffusion and subdiffusion motion depicted versus temperature $Q \propto T$ and the potential asymmetry parameter $\varphi$, respectively. For the other parameters read Fig. \ref{fig2}.}
	\label{fig9}
\end{figure}

Last but not least, we analyze impact of an initial probability distribution for the Brownian particle velocity $P(v)$ on the observed diffusion anomalies. Transient phenomena often depend on a state in which a system is initially prepared therefore it is a vital question to pose. In our case this state is described by probability distributions for the particle coordinate $x(0)$ and velocity $\dot{x}(0) = v(0)$. We focus only on the latter since the former can hardly be manipulated in many experimentally accessible physical systems that can be described by the studied model. For example, in an asymmetric SQUID device composed of three resistively and capacitively Josephson junctions the Brownian particle coordinate $x(t)$ and velocity $v(t)$ translates to the Josephson phase $\Psi(t)$ and voltage drop $V(t)$ across the setup, respectively \cite{spiechowicz2014prb}. The first quantity cannot be directly manipulated whereas the second may be easily altered by applying an external electric field. For this reason we now compare the diffusion processes occurring in the model Eq. (\ref{eq:dimlessmodel}) starting from three different types of an initial condition for the velocity $v(0)$: a Dirac delta $P(v) = \delta(v)$, a random variable distributed uniformly $P(v) = U(-2,2)$ and allocated normally with a zero mean and a standard deviation equals to one $P(v) = N(0,1)$. The result is shown in Fig. \ref{fig9}. In panel (a) we study time evolution of the diffusion coefficient $D(t)$ for selected temperature $Q = 0.001$. The discrepancies between these distributions die out after approximately $10^2$ characteristic units of time. The system relaxes to normal diffusion in a universal pattern which is independent on the form of the initial velocity distribution. It can be divided into three time domains: the early period of superdiffusion, the intermediate interval where subdiffusion is developed and approaches the asymptotic long time regime where normal diffusion occurs. Therefore at least for the studied regime the observed behaviour is omnipresent. It is confirmed in the remaining panels of this figure when we illustrate a duration of the period of superdiffusion respectively as a function of temperature $Q \propto T$ and the potential asymmetry parameter $\varphi$. In particular, the discussed extremal sensitivity to changes of the latter is robust with respect of variation of the initial velocity distribution $P(v)$.

\section{Discussion}
With this work we investigated diffusion anomalies occurring in the archetype model of an inertial Brownian ratchet. In particular, we established the clear connection between the directed transport and the transient anomalous diffusion in this setup. The key ingredients for the emergence of the latter are as follows: (i) the system is driven \emph{far away from thermal equilibrium} into a time dependent nonequilibrium state, (ii) \emph{reflection symmetry} of the potential is violated so that the directed transport may arise, (iii) \emph{ergodicity} of the corresponding deterministic system is strongly broken due to the coexistence of two counter-propagating regular attractors which almost evenly occupy the accessible phase space. The resulting dynamics is such that it is exhibiting a vanishing average nonequilibrium response implying the ballistic motion. Upon an introduction of thermal noise the system will be typically \emph{weakly nonergodic} with an extremely slow relaxation of the degrees of freedom leading to a whole range of anomalous diffusion phenomena. However, after a sufficiently long time these anomalies die out and diffusion is always normal. This is a very important conclusion since the mentioned slow relaxation may give the wrong impression that the anomalous diffusion is observed even in the asymptotic long time limit. The latter can happen only in systems whose dynamics is dominated by large and rare fluctuations that are characterized by broad distributions with power-law tails. A hallmark of these non-Gaussian distributions is the divergence of their second and/or first moment. These systems do not obey the law of large numbers and the strong convergence to the Gaussian according to the central limit theorem is broken. Consequently, diffusion anomalies may be observed even in the asymptotic long time regime \cite{kessler2010, dechant2011,zarbudaev2013,rebenshtok2014}. Despite the fact that diffusion anomalies observed in the studied model have only transient nature their duration can be fine-tuned by adjustment of the setup parameters so that they last many orders longer than characteristic time scales of the system. Therefore from an experimental point of view they may be safely treated as persistent effects.
\section{Conclusions}
In summary, we presented the detailed qualitative theory illuminating the coexistence of the directed transport as well as the anomalous diffusion processes in a generic system of an inertial Brownian ratchet. We explained the underlying physical mechanism standing behind the emergence of diffusion anomalies and the parameter-dependent control of their extended regimes. Extremely slow relaxation and ergodicity breaking typically describe glassy dynamics \cite{bouchaud1992, burov2010}. Our work shows that these features can be observed even within a more straightforward, one dimensional classical Markovian dynamics with Brownian motion of its inherent long-time Gaussian nature, i.e. without the need to introduce heavy-tailed distributions \cite{lutz2013}, nor disorder \cite{sancho2004, khoury2011, simon2013, simon2014} or many-body physics \cite{zaburdaev2015}.

The appealing strength and beauty of Brownian motion with its intrinsic Gaussian noise propagator lies in its universality and therefore our findings can be straightforwardly corroborated experimentally with a wealth of physical systems outlined in the introductory part of the article. One of the most promising setups for this purpose are optical lattices \cite{denisov2014} and asymmetric SQUID devices \cite{sterck2005, spiechowicz2014prb}. It is because of their high tunability: the period, the amplitude and the symmetry of the corresponding potential may be modified in a controlled way.

In view of the widespread applications of Brownian motor setups and ratchet devices our research may carry potential impact for further development of a working principle of a nanomotor operating on  smallest scales, such as occurring in diverse areas of nanophysics  \cite{hanggi2009,astumian2002}.
\section*{Methods}
The Fokker-Planck equation corresponding to the analyzed Langevin one Eq. (\ref{eq:dimlessmodel}) understandably cannot be solved by use of   analytical means. Therefore, in order to obtain the relevant characteristics we have to carry comprehensive numerical simulations. We did so by employing a weak version of the stochastic second-order predictor-corrector algorithm with a time-step typically set to about $(10^{-3} - 10^{-2}) \times \mathsf{T}$. Because Eq. (\ref{eq:dimlessmodel}) is a second-order differential equation, we need to specify two initial conditions, namely $x(0)$ and $\dot{x}(0)$. For some regimes the system dynamics does exhibit nonergodic behavior; thus,  in order to avoid a dependence of the presented results on a specific choice of the initial conditions, unless stated otherwise, we have chosen $x(0)$ and $\dot{x}(0)$ to be equally distributed over the intervals $[0, 2\pi]$ and $[-2,2]$, respectively. All our quantities of interest were averaged over $10^3 - 10^5$  sample trajectories. All numerical calculations have been performed by use of a CUDA environment as implemented on a modern desktop GPU. This procedure did allow for a speedup of a factor of the order $10^3$ times as compared to a common present-day CPU method \cite{spiechowicz2015cpc}.
\section*{Acknowledgments}
This work was supported in part by the MNiSW program via a Diamond Grant (J.S.).
\section*{Author Contributions}
J.S. performed all numerical simulations in this project. All authors contributed extensively to the planning, discussion of the results and writing up of this work.
\section*{Competing financial interests}
The authors declare no competing financial interests.

\end{document}